**Title: 1550 nm compatible ultrafast photoconductive material based on a GaAs/ErAs/GaAs heterostructure**

Authors: *Kedong Zhang, Yiwen Li, Yifeng Ren, Xing Fan, Chen Li, Jianfei Li, Yafei Meng, Yu Deng, Fengqiu Wang*, Hong Lu* and Yan-Feng Chen*

K. Zhang, Y. Ren, X. Fan, Prof. Y. Deng and Prof. Y. Chen
National Laboratory of Solid State Microstructures
Department of Materials Science and Engineering, College of Engineering and Applied Sciences
Nanjing University
Nanjing 210093, P. R. China
Y. Li, J. Li, Dr. Y. Meng, Prof. F. Wang
School of Electronic Science and Engineering
Nanjing University
Nanjing 210093, P. R. China
Dr. C. Li, Prof. H. Lu
National Laboratory of Solid State Microstructures
Department of Materials Science and Engineering, College of Engineering and Applied Sciences
Collaborative Innovation Center of Advanced Microstructures
Nanjing University
Nanjing 210093, P. R. China
Jiangsu Key Laboratory of Artificial Functional Materials
Nanjing 210093, P. R. China
*Corresponding author, E-mail: hlu@nju.edu.cn
                            fwang@nju.edu.cn

**Abstract**

The sub-bandgap absorption and ultrafast relaxation in a GaAs/ErAs/GaAs heterostructure are reported. The infrared absorption and 1550 nm-excited ultrafast photo-response are studied by Fourier transform infrared (FTIR) spectrometry and time-domain pump-probe technique. The two absorption peaks located at 2.0 μm (0.62 eV) and 2.7 μm (0.45 eV) are originated from the ErAs/GaAs interfacial Schottky states and sub-bandgap transition within GaAs, respectively. The photo-induced carrier lifetime, excited using 1550 nm light, is measured to be as low as 190 fs for the GaAs/ErAs/GaAs heterostructure, making it a promising material for 1550-nm-technology-compatible, high critical-breakdown-field THz devices. The relaxation mechanism is proposed and the functionality of ErAs is revealed.

**Introduction**

Terahertz technology is widely used in many fields such as security screening, biological and environmental sensing, medical diagnosis, broadband communication, and so on.[1-6] Photomixers and pulse-driven dipolar antennas fabricated from the ultrafast photoconductive materials, such as low-temperature grown GaAs (LT-GaAs), have been regarded as promising sources for coherent terahertz radiation with large tunability. [7-10] LT-GaAs has been the dominating ultrafast material since 1990s as it features all the required properties including high resistivity, high carrier mobility and short carrier lifetime.[11-14] The excess arsenic resulting from the low growth temperature forms $As_{Ga}$ antisites which create deep defect levels within the GaAs bandgap. The presence of these defect states provides an effective non-radiative relaxation pathway for the excited carriers and results in significantly shortened lifetimes. However, it sacrifices the crystalline quality so that the reliability, as the defects are hard to be well controlled. Other than LT-GaAs, GaAs heavily doped with Er (GaAs:Er) was found to be an alternative candidate to be implemented into ultrafast photoconductive devices driven by 800 nm lasers.[9, 15-16] The Er atoms can be readily and heavily incorporated into a high-quality GaAs matrix to form ErAs precipitates and these semi-metallic precipitates provide the relaxation channel for the photo-excited carriers.[9, 17-18] By suitably designing a superlattice structure, the lifetime of GaAs:Er is found to be controlled in a deterministic manner. Compared with LT-GaAs, GaAs:Er can obtain a higher crystalline quality, a higher operating temperature, and yet better transport characteristics. But foremost, the potentially highly tunable feature makes it a promising ultrafast photoconductive material, alternative to the widely used LT-GaAs.

No matter what relaxation mechanism is utilized, the GaAs-based material system needs to be excited by cross-bandgap energies, i.e. 800 nm, which are more expensive and less flexible than their 1550 nm counterparts. To simplify the pumping scheme, people turn the focus on LT-In$_{0.53}$Ga$_{0.47}$As and In$_{0.53}$Ga$_{0.47}$As:Er that can be grown on InP,[16, 19-21] because the cross-bandgap transition in In$_{0.53}$Ga$_{0.47}$As can be excited by 1550 nm. However, such change would bring problems such as lower critical breakdown fields and higher dark currents, due to the reduced bandgap. A promising solution is to use the sub-bandgap absorption of ErAs/GaAs composite material system to achieve both 1550 nm-excited ultrafast photo-response and high critical breakdown field. E. R. Brown et, al. demonstrated a GaAs-based 1550-nm continuous-wave photomixer using GaAs:Er [22-24]. Although the combination of ErAs nanoparticles and GaAs matrix has been reported, the tunability in material design is limited, for example, it is challenging to tune either the density or the size of the ErAs nanoparticles in a wide range. Therefore, the physical mechanism of the infrared absorption and ultrafast relaxation has not been confirmed yet. In particular, the functionality of ErAs itself has been neglected, mainly due to the difficulties to achieve high quality ErAs bulk films and protect it from oxidation. ErAs owns a high absorption coefficient and can be grown coherently on GaAs. How to employee this material more flexibly to achieve better performance in sub-bandgap photo-response deserves more exploration.

In this work, we design a GaAs/ErAs/GaAs heterostructure where the ErAs is sandwiched between two GaAs layers. A series of samples with different ErAs thicknesses were grown on GaAs (100) substrates by molecular beam epitaxy (MBE). The high crystalline properties of the epitaxially grown films were characterized by X-ray diffraction (XRD) and high-angle annular dark-field scanning transmission electron microscopy (HAADF-STEM). Fourier transform infrared (FTIR) spectroscopy and time-domain pump-probe technique were used to study the infrared absorption and ultrafast photo-response, respectively. Finally, we established an energy band model and proposed the energy transition mechanism for the GaAs/ErAs/GaAs heterostructure.

**Experimental**

All the heterostructure samples with different ErAs thicknesses were grown with a III-V solid-state source MBE (Veeco GENxplor) on semi-insulating GaAs (100) substrates except the Schottky junctions described below. Reflection high energy electron diffraction (RHEED) was used for in-situ monitoring of the sample surface. Elementary erbium (Er) material with a purity of 99.995% was used in a high-temperature effusion cell. All the GaAs substrates were thermally heated under an $As_2$ flux of $9 \times 10^{-6}$ torr to remove the native oxides before a 100 nm GaAs buffer layer was grown at 580 °C. Then the substrates were cooled to 500 °C for the growth of the ErAs films. Finally, a 5 nm GaAs was grown on the top for two purposes, to form a sandwich structure where the ErAs is confined, and to protect the ErAs from oxidation when exposed to air. The ErAs growth rate is 200 nm/h. The layered sample structure is shown as the insert of **Figure 1**a. The thickness of ErAs in each sample was confirmed by high resolution XRD (Bruker D8 Advanced). Cu Kα1 line is used in the high-resolution XRD and the wavelength is 0.15406 nm. The optical absorption in the range of 1.0 to 4.5 μm is measured by FTIR (Nicolet iS50). The source for the pump-probe technique is a Ti:Sapphire amplifier pumped OPA (Coherent Inc.) that outputs 1550 nm with a pulse width of ~150 fs.

**Results and discussion**

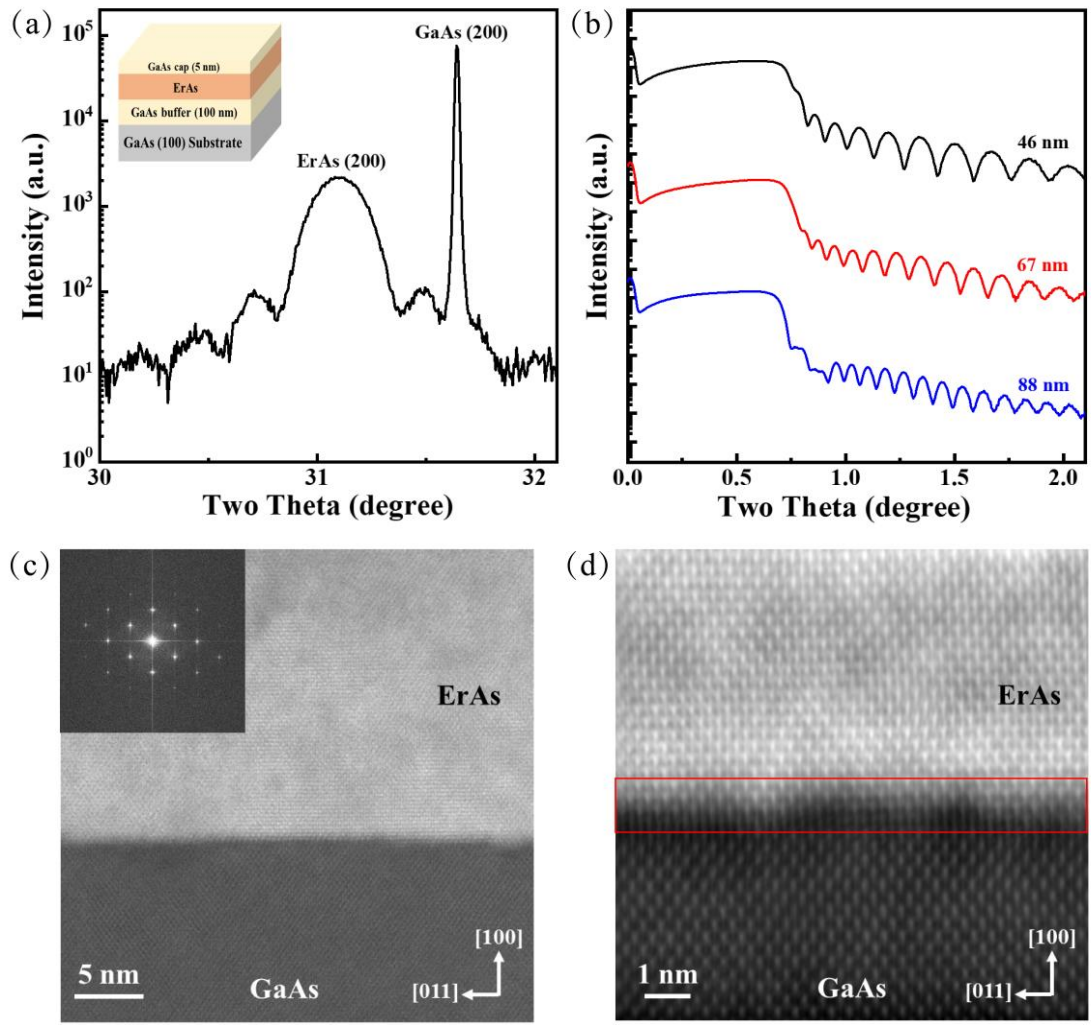

**Figure 1.** Structural characterization of the GaAs/ErAs/GaAs heterostructures grown on GaAs substrates. (a) XRD spectrum of two theta-omega scan on an epitaxially grown GaAs/ErAs/GaAs heterostructure. The layered sample structure is shown as the insert. (b) XRR spectra of three GaAs/ErAs/GaAs heterostructures with different ErAs thicknesses. And the actual ErAs thickness can be fitted using the fringes. (c) Cross-sectional HAADF STEM image of an ErAs/GaAs sample, showing the sharp interface. The inserted FFT diffraction pattern shows the coherence between the ErAs and GaAs lattices. The interfacial roughness, within 4 ML, can be seen in a STEM image with higher magnification (d).

Both the ErAs and GaAs diffraction peaks of a GaAs/ErAs/GaAs heterostructure can be observed in the XRD spectrum, as shown in Figure 1a. The two theta-omega scan was done on (200) plane to obtain the strongest diffraction signal of ErAs. The ErAs (200) diffraction peak is located at 31.1°, while the GaAs (200) diffraction peak is located at 31.6°. Thickness fringes are clearly shown around the ErAs (200) peak, indicating the high quality of the ErAs film and the ErAs/GaAs interfaces. To accurately characterize the thickness of the ErAs films, X-ray reflectivity (XRR) measurement was also carried out. The thickness can be fitted using the periodic peaks between 1° and 2° in Figure 1b, and is inversely proportional to the peak period. The ErAs thicknesses turn out to be 46, 67, 88 nm, respectively.

The ErAs/GaAs interface was studied by cross-sectional HAADF STEM on GaAs (01-1) plane, as shown in Figure 1c and 1d, to further confirm the crystalline quality of the ErAs film. The sharp interface can be seen in Figure 1c. The diffraction patterns taken from fast Fourier transformation (FFT) can be seen in the insert, indicating that the lattices of GaAs and ErAs are coherent. Although ErAs has a rocksalt structure while GaAs is zinc blende, they share the same As matrix,[25] so the As sub-lattice is continuous across the ErAs/GaAs interface. In addition, the higher quadruple symmetry of ErAs ensures its epitaxy on GaAs. The atomic resolution image is shown in Figure 1d to further confirm the interfacial quality. Although there is still some interfacial roughness which is within 4 monolayer (ML), the lattice is continuous through the interface without obvious defects.

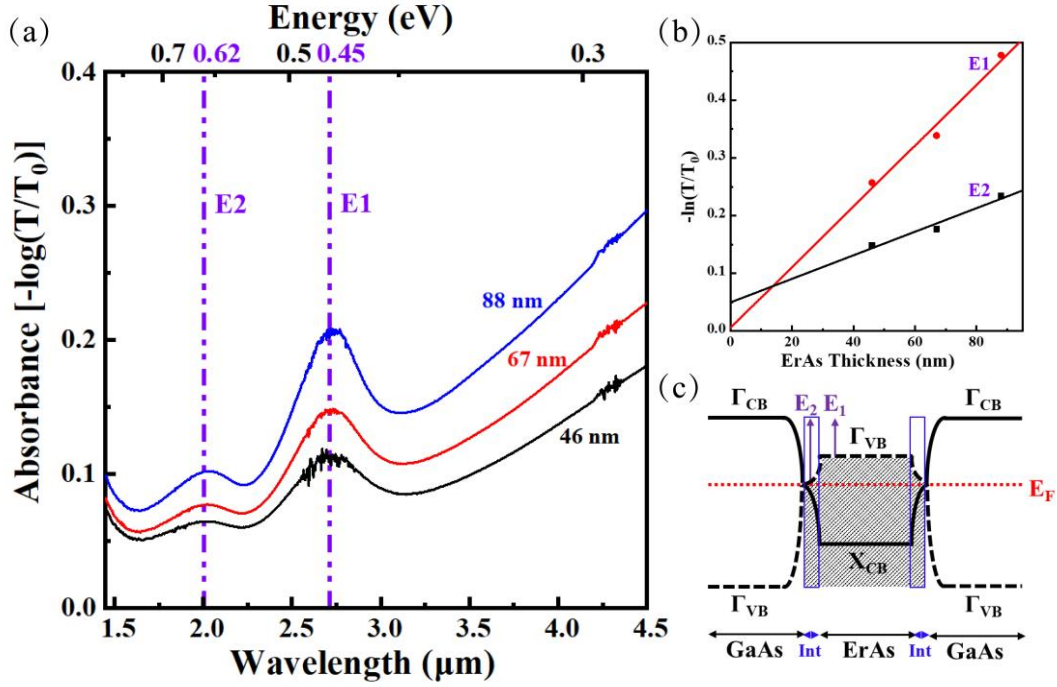

**Figure 2.** Infrared absorption of the GaAs/ErAs/GaAs heterostructures using FTIR. (a) The absorbance (calculated by –log($T/T_0$)) as a function of wavelength. (b) –ln($T/T_0$) as a function of the ErAs film thickness $t$ for peak E1 and E2, and the linear fitting to obtain the absorption coefficient (–Δ[ln($T/T_0$)]/Δ$t$). (c) Schematics of the band structure of the GaAs/ErAs/GaAs heterostructure, showing the two transitions due to ErAs.

In order to find out if there is sub-bandgap absorption in the GaAs/ErAs/GaAs heterostructure, the transmittance of all the samples were measured by FTIR from 1.0 to 4.5 μm. The corresponding absorbance was calculated by -log($T/T_0$) and plotted in **Figure 2**a, where $T$ is the measured sample transmittance, $T_0$ is the transmittance measured on a GaAs substrate. A continuous absorption is observed between 1.6 and 4.5 μm and the absorption increases as the wavelength increases. Two absorption peaks

E1 and E2 are found at 2.7 μm (0.45 eV) and 2.0 μm (0.62 eV), respectively, on the increased absorption background. And the peak intensity is proportional to the ErAs thickness. We believe that the continuous absorption is due to the semi-metallic characteristics of the ErAs layer. It is worth mentioning that the broadband absorption can be beneficial in applications using saturable absorption. The absorption below 1.5 μm is due to the detector limitation. To establish the physical origins of the two absorption features, we plot the peak intensity of E1 and E2 as a function of the ErAs thickness, as shown in Figure 2b. The absorption coefficient was approximated by $-\Delta[\ln(T/T_0)]/\Delta t$, where $t$ is the ErAs film thickness. We can obtain an absorption coefficient to be $5.26\times10^4$ cm$^{-1}$ for E1, which is 4 times higher than that reported in GaAs:Er,[26] and $2.04\times10^4$ cm$^{-1}$ for E2. For E1, the linear fitting passes the coordinate zero point, suggesting the E1 absorption is intrinsic and mainly from the ErAs itself, while the E2 absorption seems to correspond to the interaction between ErAs and GaAs. A band structure of the GaAs/ErAs/GaAs heterostructure is proposed to show the band alignments and transitions due to the experimental observations, as shown in Figure 2c.

For the ErAs films with thicknesses of tens of nanometers, the quantum confinement effect is negligible, so the absorption mainly shows the bulk nature of ErAs. The intensity of the absorption peak is proportional to the ErAs thickness while there is no shift in the peak position. It is well accepted that the conduction band minimum in ErAs is composed of the Er 5d states, while the valence band maximum is made up of the As 4p states. And the Γ-X overlap in bulk ErAs was reported to be 0.7±0.1 eV with $E_F$ being close to the midgap between the band extrema.[26] However, the band alignment between ErAs and GaAs may be determined by the Schottky states formed at the ErAs/GaAs interface. This interfacial Schottky state between GaAs and ErAs had been studied theoretically,[27-28] and confirmed using scanning tunneling spectroscopy (STS) by Palmstrøm, et al.[29] To find out this interfacial Schottky state, we grew a Schottky junction of ErAs/GaAs to study the Schottky barrier height (SBH) between ErAs and GaAs. The GaAs substrate used for the Schottky junction growth is heavily Si doped by 1×18 cm$^{-3}$, after the deoxidation, 500 nm GaAs buffer slightly doped with Si by 5×16 cm$^{-3}$ was grown at 580 °C, followed by 100 nm ErAs was grown at 500 °C to form a Schottky junction. The results show that the SBH is about 0.65 eV. We think the Fermi level is pinned by the interfacial states, so the band alignment is proposed as shown in Figure 2c. The measured E2 (0.62 eV) value is close to the SBH, so the transition may be attributed to the interfacial states. The difference may be due to a number of factors and one is that the GaAs substrate used for the Schottky junction growth is heavily doped, while the other samples were grown on semi-insulating substrates. Although the STEM images show that the epitaxially grown ErAs/GaAs is of high quality and free of interfacial defects, the pinned state may originate from how the GaAs growth is terminated and the difference in lattice structures between GaAs and ErAs. According to the band alignment, the E1 (0.45 eV) absorption corresponds to the transition between the ErAs VB and the GaAs CB at Γ point. In addition, the intensity of E1 is much stronger than E2, suggesting that E1 originates from the ErAs VB whose density of state is much higher than the interfacial Schottky states, where E2 comes from.

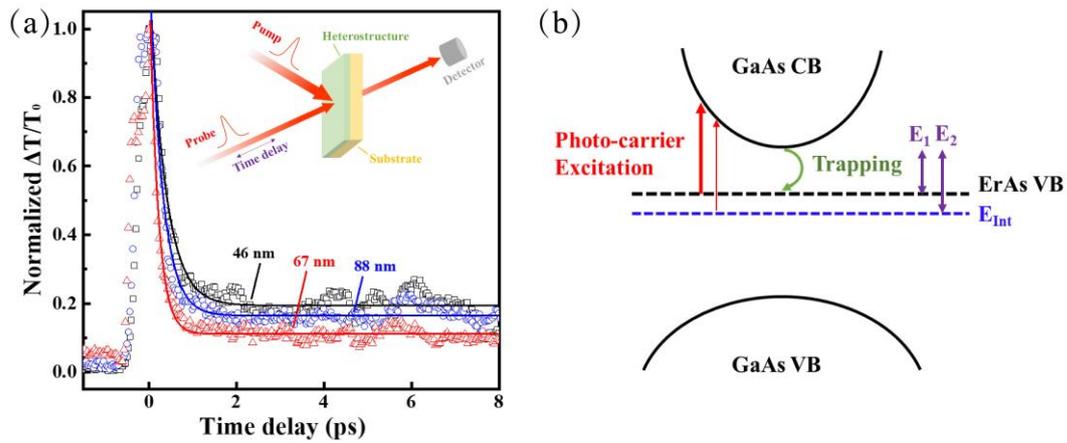

**Figure 3.** Ultrafast photo-response measurement using 1550 nm pump-probe technique. (a) Differential transmittance of the GaAs/ErAs/GaAs heterostructures. The insert is a schematic of the pump-probe measurement. (b) The band structure showing the corresponding band transitions in the process.

The time-domain pump-probe transmittance measurement was carried out to study the dynamics of carrier relaxation in the GaAs/ErAs/GaAs heterostructures. The schematic of the measurement is shown as the insert in **Figure 3**a. The light source is a Ti:Sapphire amplifier pumped OPA (Coherent Inc.) that outputs 1550 nm with a pulse width of ~150 fs. The wavelength is 1550 nm for both the pump and probe beams to investigate the sub-bandgap excitation, as GaAs is highly transparent to this wavelength. It can be seen that all the samples are characterized by a single sharp positive photo-bleaching (PB) peak followed by an exponential decay as shown in Figure 3a. A rapidly increasing $\Delta T/T$ around time zero represents a salient PB signal, which can be attributed to Pauli blocking induced phase space filling.[30-32] The following dynamics is due to the ultrafast capture process. The signal rapidly decreases by more than 80% of the peak value, which can be fitted by a single exponential $Y(t) = A\exp(-t/\tau)$ in the figure, where $\tau$ is the carrier relaxation time. The time constants for the ErAs films of 46, 67, 88 nm are 360, 190, and 290 fs, respectively, which bode well for terahertz performance.

Figure 3b shows the band transitions of the ultrafast capture process in the GaAs/ErAs/GaAs heterostructure. When the sample is excited by the 1550 nm light (0.8 eV), a significant number of carriers are excited from both the ErAs VB and the interfacial Schottky state to the GaAs CB, corresponding to the E1 (0.45 eV) and E2 (0.62 eV) transitions, respectively. The carriers from the ErAs VB dominate the excitation as proved by the relative higher intensity of the E1 absorption peak. Then the carrier can be captured through an ultrafast relaxation to the ErAs VB. The system supports ultrafast femtosecond (<400 fs) relaxation at 1550 nm is partially due to the fact that the degenerate pump-probe setup has excess energies, so that fast intra-band relaxation acts as the main relaxation channel. The appearance of a plateau-like slow relaxation may correspond to the carrier relaxation time near the GaAs band-edge. The composite of a fast femtosecond relaxation coupled with a relatively slow relaxation component is long known to benefit other ultrafast applications as well, such as a saturable absorber.[33-34] There is no obvious dependence of the carrier relaxation time on the ErAs thickness, probably because the available ErAs VB states for carrier capture is high enough and not being saturated for these thick ErAs films. More systemetic studies on thinner ErAs insertion can further reveal the relaxation mechanisms, but it is out of the scope this work.

## Conclusion

In summary, we have introduced a new structural design utilizing semi-metallic ErAs for ultrafast photo-conductive applications. A series of high quality GaAs/ErAs/GaAs heterostructures have been grown by MBE on GaAs (100) substrates. Sub-bandgap absorption has been observed in these GaAs/ErAs/GaAs heterostructures with two absorption peaks located at 2.0 and 2.7 μm, which are attributed to the transitions from the interfacial Schottky states and the ErAs VB to the GaAs CB, respectively. Ultrafast carrier relaxation (<400 fs) is revealed by the degenerate pump-probe spectroscopy operated at 1550 nm, showing the great significance in application such as cost-effective, telecom-compatible Terahertz devices.

## Acknowledgments

The authors acknowledge the support from the National Key R&D Program of China (2018YFA0306200, 2017YFA0303702, 2018YFB2200500, 2017YFA0206304), the National Natural Science Foundation of China (Grant No.51732006, No.11890702, No.51721001, NO.51702153, NO.61775093), the Natural Science Foundation of Jiangsu Province (BK20160627), the Fundamental Research Funds for the Central Universities and Jiangsu Entrepreneurship and Innovation Program. H. L. acknowledges the support from Arthur Gossard.

## References

[1]     H. H. Chen, W. L. Ma, Z. Y. Huang, Y. Zhang, Y. Huang, Y. S. Chen, *Adv. Opt. Mater.* **2019**, 7, 1801318.

[2]     B. Ferguson, X. C. Zhang, *Nat. Mater.* **2002**, 1, 26.

[3]     X. J. Fu, F. Yang, C. X. Liu, X. J. Wu, T. J. Cui, *Adv. Opt. Mater.* **2020**, 8, 1900628.

[4]     R. R. Hartmann, J. Kono, M. E. Portnoi, *Nanotechnology* **2014**, 25, 322001.

[5]     M. Lee, M. C. Wanke, *Science* **2007**, 316, 64.

[6]     P. H. Siegel, *IEEE Trans. Microw. Theory Tech.* **2004**, 52, 2438.

[7]     R. Mendis, C. Sydlo, J. Sigmund, M. Feiginov, P. Meissner, H. L. Hartnagel, *Int. J. Infrared Milli. Waves* **2005**, 26, 201.

[8]     S. Verghese, K. A. McIntosh, S. Calawa, W. F. Dinatale, E. K. Duerr, K. A. Molvar, *Appl. Phys. Lett.* **1998**, 73, 3824.

[9]     C. Kadow, A. W. Jackson, A. C. Gossard, S. Matsuura, G. A. Blake, *Appl. Phys. Lett.* **2000**, 76, 3510.

[10]    I. D. Henning, M. J. Adams, Y. Sun, D. G. Moodie, D. C. Rogers, P. J. Cannard, S. J. Dosanjh, M. Skuse, R. J. Firth, *IEEE J. Quantum Electron.* **2010**, 46, 1498.

[11]    A. Hache, Y. Kostoulas, R. Atanasov, J. L. P. Hughes, J. E. Sipe, H. M. vanDriel, *Phys. Rev. Lett.* **1997**, 78, 306.

[12]    B. Heshmat, H. Pahlevaninezhad, Y. Pang, M. Masnadi-Shirazi, R. B. Lewis, T. Tiedje, R. Gordon, T. E. Darcie, *Nano Lett.* **2012**, 12, 6255.

[13]    X. Liu, A. Prasad, W. M. Chen, A. Kurpiewski, A. Stoschek, Z. Lilientalweber,

E. R. Weber, *Appl. Phys. Lett.* **1994**, 65, 3002.

[14]  M. Tani, S. Matsuura, K. Sakai, S. Nakashima, *Appl. Opt.* **1997**, 36, 7853.

[15]  S. N. Gilbert Corder, J. K. Kawasaki, C. J. Palmstrøm, H. T. Krzyzanowska, N. H. Tolk, *Phys. Rev. B* **2015**, 92, 134303.

[16]  M. Griebel, J. H. Smet, D. C. Driscoll, J. Kuhl, C. A. Diez, N. Freytag, C. Kadow, A. C. Gossard, K. Von Klitzing, *Nat. Mater.* **2003**, 2, 122.

[17]  A. K. Azad, R. P. Prasankumar, D. Talbayev, A. J. Taylor, R. D. Averitt, J. M. O. Zide, H. Lu, A. C. Gossard, J. F. O'Hara, *Appl. Phys. Lett.* **2008**, 93.

[18]  K. Zhang, S. Xia, C. Li, J. Pan, Y. Ding, M.-H. Lu, H. Lu, Y.-F. Chen, *Adv. Mater. Interfaces* **2020**, 7.

[19]  I. Kostakis, D. Saeedkia, M. Missous, *IEEE Trans. THz Sci. Technol.* **2012**, 2, 617.

[20]  B. Sartorius, M. Schlak, D. Stanze, H. Roehle, H. Kuenzel, D. Schmidt, H. G. Bach, R. Kunkel, M. Schell, *Opt. Express* **2009**, 17, 15001.

[21]  D. Stanze, A. Deninger, A. Roggenbuck, S. Schindler, M. Schlak, B. Sartorius, *J. Infrared Milli. Terahz. Waves* **2011**, 32, 225.

[22]  J. R. Middendorf, E. R. Brown, *Opt. Express* **2012**, 20, 16504.

[23]  W. D. Zhang, J. R. Middendorf, E. R. Brown, *Appl. Phys. Lett.* **2015**, 106, 021119.

[24]  E. R. Brown, A. Mingardi, W. D. Zhang, A. D. Feldman, T. E. Harvey, R. P. Mirin, *Appl. Phys. Lett.* **2017**, 111, 031104.

[25]  D. O. Klenov, J. M. O. Zide, J. D. Zimmerman, A. C. Gossard, S. Stemmer,

*Appl. Phys. Lett.* **2005**, 86, 241901.

[26] M. A. Scarpulla, J. M. O. Zide, J. M. LeBeau, C. G. Van de Walle, A. C. Gossard, K. T. Delaney, *Appl. Phys. Lett.* **2008**, 92, 173116.

[27] W. R. L. Lambrecht, A. G. Petukhov, B. T. Hemmelman, *Solid State Commun.* **1998**, 108, 361.

[28] K. T. Delaney, N. A. Spaldin, C. G. Van de Walle, *Phys. Rev. B* **2010**, 81, 165312.

[29] J. K. Kawasaki, R. Timm, K. T. Delaney, E. Lundgren, A. Mikkelsen, C. J. Palmstrøm, *Phys. Rev. Lett.* **2011**, 107, 036806.

[30] C. Zhu, F. Wang, Y. Meng, X. Yuan, F. Xiu, H. Luo, Y. Wang, J. Li, X. Lv, L. He, Y. Xu, J. Liu, C. Zhang, Y. Shi, R. Zhang, S. Zhu, *Nat. Commun.* **2017**, 8, 14111.

[31] S.-H. Cho, W.-S. Chang, J.-G. Kim, K.-H. Whang, K.-S. Choi, S.-H. Sohn, *Appl. Surf. Sci.* **2008**, 254, 3370.

[32] D. Swain, P. T. Anusha, T. S. Prashant, S. P. Tewari, T. Sarma, P. K. Panda, S. V. Rao, *Appl. Phys. Lett.* **2012**, 100, 141109.

[33] Y. Sun, Y. Meng, R. Dai, Y. Yang, Y. Xu, S. Ziu, Y. Shi, F. Xiu, F. Wang, *Opt. Lett.* **2019**, 44, 4103.

[34] Y. Meng, C. Zhu, Y. Li, X. Yuan, F. Xiu, Y. Shi, Y. Xu, F. Wang, *Opt. Lett.* **2018**, 43, 1503.